# CRUC: Cold-start Recommendations Using Collaborative Filtering in Internet of Things

Daqiang Zhang[a,*], Qin Zou[b], Haoyi Xiong[c]

[a] *School of Computer Science, Nanjing Normal University, Nanjing, Jiangsu Province, 210046,China*
[b] *School of Remote Sensing and Information Engineering, Wuhan University, Wuhan, Hubei Province, 430072, China*
[c] *Department of Telecommunication, Institute Telecome – Telecom SudParis, Evry, 91011, France*

**Abstract**

The Internet of Things (IoT) aims at interconnecting everyday objects (including both things and users) and then using this connection information to provide customized user services. However, IoT does not work in its initial stages without adequate acquisition of user preferences. This is caused by cold-start problem that is a situation where only few users are interconnected. To this end, we propose CRUC scheme --- Cold-start Recommendations Using Collaborative Filtering in IoT, involving formulation, filtering and prediction steps. Extensive experiments over real cases and simulation have been performed to evaluate the performance of CRUC scheme. Experimental results show that CRUC efficiently solves the cold-start problem in IoT.

*Keywords:* Cold-start Problem, Internet of Things, Collaborative Filtering

## 1. Introduction

The Internet of Things (IoT) refers to a self-configuring network in which everyday objects are interconnected to the Internet [1] [2]. IoT deploys sensors in infrastructures (e.g., rooms and buildings) to get a heightened awareness of real-time events. It also employs sensors capturing contextual information about objects (e.g., user preferences) to achieve an enhanced situational awareness [3] [21-26]. Readings from a large number of sensors for various objects are enormous, but only a few of them are useful for a specific user. Thus, IoT customizes services to users according to their preferences or behaviors that are acquired by sensors.

However, IoT is seriously limited by a cold-start problem that IoT cannot draw any inferences for users about which it has not yet collected sufficient information. The cold-start problem usually occurs at the beginning of constructing IoT systems or new objects or users are tracked and reflected in the IoT network. The cold-start problem is fundamental in IoT, which results in inaccurate recommendations and long latency in response to recommendation requests. Suppose that IoT employs Radio Frequency IDentification (RFID) [20] to keep track of user locations and then provides a tourism navigation service. Sometimes, IoT is supposed to support longer-range, more complex tourism planning and decision making. For example, IoT makes a travel plan for the ***May Day Holiday***. But without previous information of user location and behaviors, IoT fails to be deep situation awareness.

In general, the cold-start problem is characterized by two features --- scalability and sparsity. The former feature is caused by the number of tracked objects and their interaction data. IoT uses sensors to track a huge number of objects, as well as their interaction. Thus, IoT gathers an extra-large data of objects' interaction. This incurs that IoT falls short of quickly responding to requesters. The latter feature arises from the information density. Although IoT gets a large amount of sensor readings about objects, it acquires little information about a specific object, particularly in a cold-start situation. Correspondingly, it is a non-trivial job to predict user preference on the basis of cold-start IoT systems.

To this end, we conduct the first research on investigating the cold-start problem in IoT. In this paper, we propose a way to acquire such user information by Collaborative Filtering (CF), which is referred to as CRUC --- Cold-start Recommendations Using Collaborative Filtering in Internet of Things. CRUC consists of formulation, filtering and prediction steps. In the beginning, CRUC formulates the cold-start problem in IoT as a surmise of user preferences. Then, it picks the most important users and their information by introducing significant users. Finally, CRUC comes up with a fusion policy to predict user preferences.

The rest of this paper is organized as follows. Section II gives a preliminary introduction to our work. Section III discusses the proposed scheme CRUC in detail. Section IV reports the experimental results and Section V concludes our work.

## 2. Preliminary

In this section, we firstly overview the Collaborative Filtering (CF). Then, we introduce our notations to facilitate understanding of our work.

### 2.1. Collaborative Filtering

Collaborative Filtering (CF) is an enabling technique that allows recommender systems to recommending products or services that they are likely to be of interest to users. CF has been widely adopted as a salient part in Amazon [4], Google [5], Netflix [6] and Yahoo [7].

Broadly, CF consists of two primary categories — memory-based and model-based schemes. Memory-based schemes identify like-minded users or similar items over the entire item-user matrix. As a result, they often achieve high levels of accuracy as well as poor scalability [8] [9]. Memory-based schemes can be further classified into item-based and user-based schemes. In contrast, model-based schemes narrow down the searching scope of like-minded users or similar items by exploiting models mainly from machine learning and artificial intelligence [10] [11].

### 2.2. Notations

In order to keep readability of the following parts in this paper, we firstly introduce the related notations. Table 1 illustrates the notations in the design of the proposed scheme.

Table 1. Notations in the design of the proposed scheme

| Notations | Description |
|---|---|
| I | The set of items |
| U | The set of users |
| \|I\| | The number of items in the item-user matrix |
| \|U\| | The number of users in the item-user matrix |
| $I_u$ | The set of items rated by the user *u* |
| $U_i$ | the set of users who have rated the item *i* |
| $r_{u,i}$ | the rating that the user u rates the item *i* |
| $\overline{r_i}$ | the average rating of the item *i* |
| $\overline{r_u}$ | the average rating of the user *u* |
| $P_{u,i}$ | the predicted rating on the item *i* by the user *u* |
| $S_i$ | the set of the item *i*'s similar items |
| $S_u$ | the set of the user *u*'s like-minded users |

## 2.3. Item-based and User-based CF Schemes

Item-based CF regards that a user may prefer similar items. Given an active *u* and an active item *i*, Eq. 1 shows the manner that item-based CF work, where *sim(i, j)* is the similarity measurement between items *i* and *j*.

$$P_{u,i} = \overline{r_i} + \frac{\sum_{j \in I_u} sim(i,j) \bullet (r_{u,j} - \overline{r_j})}{\sum_{j \in I_u} sim(i,j)} \quad (1)$$

In contrast, user-based CF assumes that like-minded user may like the same item. Given two active users *u* and *v*, Eq. 2 displays the way that user-based CF work, where *sim(u,v)* is the similarity between active users.

$$P_{u,i} = \overline{r_u} + \frac{\sum_{v \in U_i} sim(u,v) \bullet (r_{v,i} - \overline{r_v})}{\sum_{v \in U_i} sim(u,v)} \quad (2)$$

Note that the similarity measurement for items and users is usually computed using Pearson Coefficient Correlation (PCC) or Vector Space Similarity (VSS) algorithms [8] [9] [12] [13].

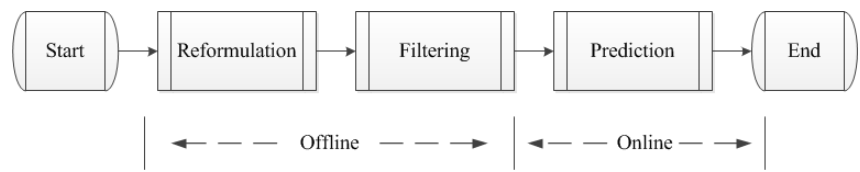

Figure 1: The CRUC scheme

In this paper, we use PCC to compute similarity of items and users, which is defined as the covariance of the two variables (e.g., X, Y) divided by the product of their standard deviations. PCC is shown in Equation 3, where *ux* and *uy* are mean values of variables *X* and *Y*, and $\sigma_x$ and $\sigma_y$ are standard deviation of variables X and Y.

$$sim(X,Y) = \frac{E[(X-ux)\bullet(Y-uy)]}{\sigma_x \sigma_y} \quad (3)$$

## 3. CRUC: Cold-start Recommendations Using Collaborative Filtering in Internet of Things

IoT involves a large number of sensors, as well as their readings for objects. In the initial phase, IoT deploys sensors into infrastructure for the purpose of tracking objects. Although the number of sensor readings is enormous, there are only a few readings for a specific user is few. In this case, IoT cannot provide any personalized services to users. This is the cold-start situation. Note that we use objects and users alternatively in the contexts without ambiguity.

To this end, we propose in this section the CRUC scheme — Cold-start Recommendations Using Collaborative Filtering in Internet of Things. In the following section, we overview it before discussing it in detail.

*3.1. Overview of CRUC Scheme*

Figure 1 illustrates the CRUC scheme, consisting of offline and online phases. In the offline phase, CRUC involves two steps — reformulation and filtering. In the online phase, CRUC contains one step — prediction. We intentionally design CRUC to two phases, and move the compute-intensive operations in the offline phase. This approach considerably reduces the computation overhead in the online phase and accelerates the response.

In the reformulation step, we handle the cold-start problem in IoT from a new perspective. We suppose that users are tracked by sensors, and items are sensor readings that are acquired in the process of sensing users. Given an active user *u*, its location is regarded as items. User u may stay in many places and thus generates different location corresponding to every place. For every place, we denote that the probability that user u stays in a location as a rating (e.g., the rating for living bedrooms of user u may be the highest score). By recording all users and their location information, we get a rating matrix as well as their ratings for a specific location. We go further by normalizing the ratings. Thus, we construct a standardized item-user matrix. This matrix is characterized by sparsity owing to the few ratings, and scalability owing to the large-scale of the item-user matrix.

We get findings from experiments that not all ratings positively contribute to the prediction of user preferences. Low-level ratings reflects the user dislike toward certain items, which negatively affect the recommendation performance. In fact, only the most significant users and their ratings can be used for recommendation. We introduce significant users to identify users who have a significant influence on the recommendation. Then, we cluster these users into different clusters. Consider that the rating density of the matrix composed of the significant users is still sparse. We introduce a smoothing policy to minimize the sparsity influence.

In the last step, we predict user preferences. We propose a fusion policy to fuse three sources of ratings in the recommendation — the ratings given to similar items from the same user, to the same item from like-minded users, and given to similar items from like-minded users. Let *M* be the number of similar items, *K* be the number of like-minded users, *SIR* be the ratings by item-based prediction, *SUR* be the ratings by user-based prediction, *SUIR* be the ratings by hybrid prediction of item-based and user-based prediction with fusion parameters $\lambda$ and $\delta$, and *SR* be the ratings by *CRUC* scheme.

### 3.2. Identifying Significant Users

Let $\rho_u$ = |Iu|/|I|, Definition 1 defines the significant users. By specifying the threshold $\overline{\rho_u}$, CRUC selects the most important users.

**Definition 1.** Frequent raters are users whose rating density $\rho_u$ is bigger than the threshold $\overline{\rho_u}$, i.e., $\rho_u > \overline{\rho_u}$, where is the average rating density of all users.

### 3.3. Clustering significant users

We cluster significant users due to two reasons. One is to remove the diversity in user ratings; the other is to alleviate the influence of rating sparsity.

We use K-means method to cluster significant users. Thus, we classify users into user clusters. The K-means method trains the data iteratively and assigns every user to a cluster whose centroid is the closest to him or her [8] [9] [12].

### 3.4. Smoothing user ratings

Sensor readings are noisy and can easily be error-prone [14] [15]. Thus, the ratings based on sensor readings are not justified. Therefore, we come up a smooth policy.

The missing values in every user cluster are smoothed by the rating deviation of the user cluster, given as Eq. 4, where $UC_u$ is the user cluster that includes the user $u$

$$r_{u,i} = \overline{r_u} + \sum_{u' \in UC_u} (r_{u',i} - \overline{r_{u'}}) / |UC_u| \qquad (4)$$

### 3.5. Fusing predictions

When IoT predicts user preferences, it has three kinds of ratings — the ratings from the same user made on the similar items, like-minded users made on the same item and like-minded users made on the similar items. Given that these ratings affect the recommendation diversely, we share a fusion policy proposed in our previous work [12]. *CRUC* chooses *SUR* as the main prediction source, and SIR and *SUIR* as supplementary sources. Let $\lambda$ and $\delta$ be the fusion parameters whose values are between 0 and 1, Equation 5 defines the fusion policy.

$$SR : \overline{r_{u_b,i_a}} = (1-\delta) \bullet (1-\lambda) \bullet SIR + (1-\delta) \bullet \lambda \bullet SUR + \delta \bullet SUIR \qquad (5)$$

## 4. Experiments

We conduct a series of experiments including both simulation and real implementation to evaluate the proposed scheme. We report the experimental results in our simulation. We select the MAE (Mean Absolute Error) and RMSE (Root Mean Squared Error) that can be regarded canonical measurement in CF (e.g., in [8] [9] [16] [17] [18]).

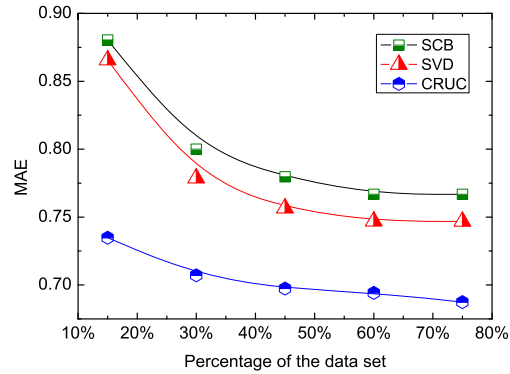

Figure 2: MAE over the MovieLens data set (A small value means a better performance)

We use MovieLens data (http://www.grouplens.org/) set to simulate the ratings in the cold-start IoT system. This data set is generated by an online recommender system since 1992. We ran the program on a cluster of Ubuntu 64-bit OS that involves 4 computers. Every computer is with 8 GB RAM and Intel Xeon E5405 2.00 GHz dual CPUs. Table 2: Statistics of MovieLens dataset

| **Features** | **MovieLens** | **Features** | **MovieLens** |
| --- | --- | --- | --- |
| Number of users | 71567 | Global avg. rating | 3.5124 |
| Number of items | 10681 | Avg. #of rated items per user | 139.7 |
| Number of ratings | 10000054 | Avg. #of rated users per item | 936.2 |
| Data density | 1.31% | | |

*4.1. Estimated accuracy*

We select the 15%, 30%, 45%, 60% and 70% of the data set as the training sets, and select the remaining part 15% of the data set as the test set. We use cross validation to evaluate the proposed schemes. We select the SCB [8] and SVD [19] as benchmark schemes. We set the fusion parameters $\lambda$ and $\delta$ (refer to [12] for detail) as 0.75 and 0.1, correspondingly.

Figure 2 and Figure 3 illustrate the estimated accuracy over the MovieLens data set. With the increase of the size of the training set (i.e., from 15% to 75%), the MAEs of all schemes show a downward trend. When the size percentage of the training set is between 40% to 70%, the MAEs of all schemes keep at a lower value, indicating that these schemes achieve the better estimated accuracy. This is because that they exploit more and more useful ratings in the item-user matrix. When the size percentage of the training set is bigger than 70%, the MAEs of all schemes do not change considerably. This indicates that 40% up to 70% percentage of the data set can represent the entire data set in prediction.

On the other side, we find that CRUC scheme gets the lowest MAE among all schemes, implying that CRUC outperforms SVD and SCB schemes. This is mainly because three reasons. One is that CRUC identifies the significant users. Moreover, CRUC removes the user rating diversity by a smooth policy within every user cluster. The last reason is that CRUC efficiently makes use of diverse rating sources by applying a fusion policy.

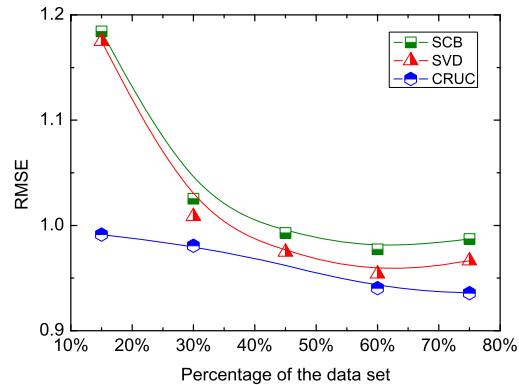

Figure 3:RMSE over the MovieLens data set (A small value means a better performance)

Note that CRUC method involves several parameters, including the threshold for identifying significant users, the fusion parameter for fusing recommendation from item-based and user-based schemes. Owing to the limit of the page space, we omit this part.

## 5. Conclusions

In this paper, we have studied the cold-start problem in the Internet of Thing (IoT). We have proposed CRUC scheme -— Cold-start Recommendations Using Collaborative Filtering in IoT, containing offline and online phases, i.e., the reformulation, filtering and prediction steps. Experimental results show that CRUC can efficiently solve the cold-start problem.

However, CRUC could be improved in several aspects. Firstly, we need to finish it in the real implementation on the basis of RFID-tracked systems. Then, we will take advantage of more aspects of the data, e.g., dates associated with the ratings and attributes of items and users, which may reflect the changes of user preferences.

**Acknowledgements**

This work is supported by the National Natural Science Foundation of China (Grant Nos. 61103185, 61003247 and 61073118), the Start-up Foundation of Nanjing Normal University (Grant No. 2011119XGQ0072), and Natural Science Foundation of the Higher Education Institutions of Jiangsu Province, China (Grant No. 11KJB520009). This work is also supported by Major Program of National Natural Science Foundation of Jiangsu Province (Grant No. BK211005).